\documentstyle[prl,aps,preprint]{revtex}
\newtheorem{theorem}{Theorem}

\newtheorem{corollary}[theorem]{Corollary}

\begin{document}
\draft


\title{
Information-Disturbance Theorem for Mutually Unbiased Observables
}

\author{Takayuki Miyadera$\ ^*$
and Hideki Imai$\ ^{*,\dagger}$
}

\address{$\ ^*$
Research Center for Information Security (RCIS),\\
National Institute of Advanced Industrial 
Science and Technology (AIST).\\
Daibiru building 1102,
Sotokanda, Chiyoda-ku, Tokyo, 101-0021, Japan.
\\
(e-mail: miyadera-takayuki@aist.go.jp)
\\
$\ ^{\dagger}$
Institute of Industrial Science, 
University of Tokyo. \\
4-6-1 Komaba, Meguro-ku, Tokyo 153-8505, Japan .
}

\maketitle
\begin{abstract}
We derive a novel version of information-disturbance 
 theorems for mutually unbiased observables.
We show that the information gain 
by Eve inevitably makes the outcomes 
by Bob in the conjugate basis not only 
erroneous but random. 
\end{abstract}

\section{Introduction}
 In 1984, Bennett and Brassard\cite{BB84} proposed a 
quantum key distribution protocol which is now called 
as BB84 protocol. Its unconditional security 
was first proved by Mayers\cite{Mayers} in 1996 and after his proof 
the various proofs\cite{HKL,Biham,Shor} have appeared.
Among them, a proof by Biham, Boyer, Boykin, Mor, and 
Roychowdhury\cite{Biham}
is based upon 
a so-called {\it information disturbance  theorem}.
According to the theorem, the information gain by 
Eve inevitably induces {\it errors} in outcomes obtained by Bob.
This disturbance enables Alice and Bob to notice the 
existence of eavesdroppers.
As well as its application to 
BB84 protocol, since it can be regarded as an 
information theoretic version of uncertainty relation,
the theorem has attracted many authors\cite{FuchsPeres,Fuchs,Winter}.
Recently, Boykin and Roychowdhury\cite{Boykin} showed a 
simple proof of the theorem in an arbitrary dimension 
by using purification technique 
and trace norm inequality. We, in this paper, 
derive a different version of the theorem. Our information-disturbance 
 theorem is an inequality between the information gain 
by Eve and the {\it randomness} (rather than 
error probability) of the outcomes obtained by Bob.
We compare our theorem with the previous one and 
discuss its implication.
\section{Setting}
Let us begin with a setting. 
Three characters, Alice, Bob and Eve 
play their roles. Our setting is a simplified version of 
BB84 quantum key distribution protocol.
The following analysis, however, can be 
applied to the full BB84 protocol with 
public discussion 
procedures.
Let us consider two pairs of orthogonal states,
$b:=\{|0\rangle, |1\rangle\}$ and its {\it conjugate} 
$\overline{b}:=
\{|\overline{0}\rangle,|\overline{1}\rangle\}$
in ${\bf C}^2$. They are assumed mutually unbiased with 
each other. That is,
\begin{eqnarray*}
\langle i| \overline{k} \rangle =\sqrt{\frac{1}{2}}(-1)^{ik}
\end{eqnarray*}
holds for each pair of $i,k \in \{0,1\}$.
Alice first selects $b$ or $\overline{b}$ which is used to 
encode a random number. 
Alice next randomly generates an $N$-bits sequence $i\in \{0,1\}^N$ with 
probability $p(i)=\frac{1}{2^N}$. 
We write $A$ a random variable representing this $N$-bits sequence.
Alice encodes this information on $N$-qubits and sends them to Bob.
For instance, suppose that Alice selects $b$ and generates a 
sequence $i=i_1 i_2\cdots i_N$, she sends the corresponding state $|i\rangle
=|i_1\rangle \otimes |i_2\rangle
 \otimes \cdots \otimes |i_N\rangle \in {\bf C}^2 
\otimes \cdots \otimes {\bf C}^2=:{\cal H}_A\simeq {\cal H}_B$ to Bob.
If the conjugate
 basis $\overline{b}$ and a sequence $j=j_1 j_2 \cdots j_N$ are 
chosen, 
the state sent to Bob is $|\overline{j}\rangle=
|\overline{j_1}\rangle \otimes |\overline{j_2}\rangle 
\otimes \cdots \otimes |\overline{j_N}\rangle
\in {\cal H}_A$.
Alice, after confirming that Bob actually 
has received $N$-qubits, informs him 
of the basis she used. 
Bob makes a measurement with respect to 
the basis and obtains an outcome. Let us 
write $B$ the random variable representing 
this outcome. If there is no eavesdropper, 
$A=B$ naturally follows.
Eve wants to obtain the information of the random variable $A$.
For the purpose, Eve prepares an apparatus and makes it interact with 
the $N$-qubits sent to Bob by Alice.  
Let us denote ${\cal H}_E$ a Hilbert space describing Eve's apparatus.
In general, 
Eve's operation is described by a unitary operator $U$,
\begin{eqnarray}
U:{\cal H}_E \otimes {\cal H}_A &\to& {\cal H}_E \otimes {\cal H}_B 
\nonumber \\  
|0\rangle \otimes |i\rangle &\mapsto& \sum_j |E_{ij}\rangle \otimes |j\rangle,
\label{Uinb}
\end{eqnarray}
where $|0\rangle$ is a normalized vector in 
${\cal H}_E$ and $\{|E_{ij}\rangle\}\subset {\cal H}_E$ 
satisfies unitarity condition: 
$\sum_{j\in \{0,1\}^N}
 \langle E_{ij}|E_{kj}\rangle
=\delta_{ik}$.
After this interaction, 
Eve tries to make an optimal
 measurement on her apparatus to 
 extract the information of $A$.
\section{Information-Disturbance  Theorem}
\subsection{Information v.s. Error}
One can show that 
if Eve's operation yields herself to gain 
large information, error probability in qubits sent to Bob 
in the conjugate basis becomes inevitably large. It has been
 called as
{\it information-disturbance  theorem} and 
was proved in \cite{Biham,Boykin}. 
\par
The representation (\ref{Uinb}) depends upon the 
choice of the basis.
It is useful to rewrite the same unitary operator
in the conjugate basis, $\overline{b}$.
Using $|\overline{i}\rangle =
\sum_{l\in \{0,1\}^N} |l\rangle \langle l|\overline{i}\rangle$ and
$|i\rangle =\sum_{j\in \{0,1\}^N} |\overline{j}\rangle \langle 
\overline{j}|i\rangle$, we obtain
$
U|0\rangle \otimes |\overline{l}\rangle
=
\sum_{s\in \{0,1\}^N} |\overline{E}_{ls}\rangle\otimes |\overline{s}\rangle,
$
where
$|\overline{E}_{ls}\rangle
:=
\sum_{i,j\in \{0,1\}^N}|E_{ij}\rangle\langle \overline{s}|j\rangle
\langle i|\overline{l}\rangle.
$
\par
When Alice chooses basis $b$ and a sequence $i\in \{0,1\}^N$,
a state obtained by Eve is computed as 
$
\rho_{Eve}^i:=\sum_{j\in \{0,1\}^N} |E_{ij}\rangle\langle E_{ij}|.
$
Later we consider how much information Eve can extract from 
it. 
When Alice chooses another basis $\overline{b}$ and a 
sequence $i$, Bob obtains a state 
\begin{eqnarray}
\overline{\rho}^i_{Bob}=\sum_{j,l\in \{0,1\}^N}\langle \overline{E}_{il}
|\overline{E}_{ij}\rangle |\overline{j}\rangle 
\langle \overline{l}|
\label{Bobstate}
\end{eqnarray}
in the presence of Eve. Later we consider 
the error it induces to the outcome.
\par
Let us begin with Eve's information gain.
Eve performs a measurement (POVM) $X:=\{X_{\alpha}\}$ on 
her state. (POVM is a family of positive operators satisfying
$\sum_{\alpha}X_{\alpha}={\bf 1}$.) We put $E[X]$ 
a random variable representing the outcome.
Probability to obtain an outcome $\alpha$ is 
$p(\alpha|i,b)=\mbox{tr}\left(X_{\alpha}\rho_{Eve}^i\right)$.
Information gain 
by Eve with respect to a POVM $X$ is 
calculated as,
\begin{eqnarray*}
I\left(A:E[X]|b\right)
&=&
H\left(A|b\right)+H\left(E[X]|b\right)-
H\left(A,E[X]|b\right)
\\
&=&
\frac{1}{2^N}
\sum_{\alpha}\sum_{i}
p(\alpha|i)
\left( \log p(\alpha|i)
-\log \sum_j p(\alpha|j)\right) +N,
\end{eqnarray*}
where $H(\cdot)$ means Shannon entropy.
What we are interested in is its optimal value with respect to 
all the possible measurements by Eve:
\begin{eqnarray*}
I\left(A:E|b\right):=
\sup\left\{I\left(A:E[X]|b\right)|
X=\{X_{\alpha}\}\mbox{is a POVM in }{\cal H}_E \right\}.
\end{eqnarray*}
\par
Now we consider outcomes obtained by Bob in the conjugate basis.
Remind that when Alice chooses basis $\overline{b}$, 
the state sent to Bob is (\ref{Bobstate}).
Bob makes a measurement of an observable $\sum j|\overline{j}\rangle
\langle \overline{j}|$.
We put $B$ a random variable for 
this outcome.
The probability to obtain each outcome is expressed as
$
p(j|i,\overline{b})=\langle \overline{E}_{ij}|\overline{E}_{ij}\rangle.
$
Thus probability to obtain an outcome whose difference from input is 
$c\in \{0,1\}^N$, is 
\begin{eqnarray}
p(B=A\oplus c|\overline{b})&:=&
\sum_i \frac{1}{2^N}p(i\oplus c|i,\overline{b})\nonumber \\
&=&\frac{1}{2^N}
\sum_i \langle \overline{E}_{i\ i\oplus c}|\overline{E}_{i\ i\oplus c}\rangle,
\label{defofP1}
\end{eqnarray}
where the symbol ``$\oplus$" is
a bit-wise XOR operation.
By use of these quantities, the information-disturbance  theorem
obtained by Boykin and Roychowdhury is expressed as\cite{Note}
\begin{eqnarray}
I(A:E|b)\leq 4N \sqrt{\sum_{c \neq 0}p(B=A\oplus c|\overline{b})},
\label{Boykins}
\end{eqnarray}
whose right hand side is proportional to the square root of the 
error probability in Bob's outcome. 
That is, their theorem claims that 
the information gain by Eve makes 
Bob's outcome in conjugate basis {\it erroneous}.
\subsection{Information v.s. Randomness}
We next derive a new information-disturbance 
theorem which relates information gain by Eve with 
randomness in Bob's outcome.
\par
To estimate the information gain 
by Eve, we introduce a 
symmetrized attack as in \cite{Biham}. 
We add $N$ auxiliary qubits to Eve's apparatus and 
thus the Eve's Hilbert space is dilated to 
${\cal H}_{E'}
:={\bf C}^2 \otimes \cdots \otimes {\bf C}^2 \otimes {\cal H}_E$. 
Introduce a set of new vectors $\{|E^s_{ij}\rangle\}$ in this Hilbert space 
${\cal H}_{E'}$ as
\begin{eqnarray*}
|E^s_{ij}\rangle:
=\sqrt{\frac{1}{2^N}}\sum_{m\in \{0,1\}^N}
(-1)^{m\cdot (i\oplus j)}|m\rangle \otimes |E_{i\oplus m\ j\oplus m} \rangle,
\end{eqnarray*} 
where ``$\oplus$" is again a bit-wise XOR operation and 
``$\cdot$" represents bit-wise multiplications
followed by their summation. 
Introduce a {\it symmetrized} attack as
\begin{eqnarray*}
U^s:{\cal H}_{E'} \otimes {\cal H}_A &\to&
{\cal H}_{E'} \otimes {\cal H}_B
\\
(|0\rangle \otimes |0\rangle)\otimes |i\rangle 
&\mapsto& 
\sum_j |E^s_{ij}\rangle \otimes |j\rangle
\end{eqnarray*}
which can be extended to satisfy unitarity condition \cite{Biham}. 
Although this symmetrized attack is different from 
the original attack, it is shown below that to treat 
this new attack is useful.
\par
If we employ the symmetrized attack,  Eve has a state described as
$
\rho^{i}_{Eve,sym}:=\sum_{j\in \{0,1\}^N}|E^s_{ij}\rangle
\langle E^s_{ij}|.
$
To extract the
information from it,
she can measure the value of the auxiliary $N$-qubits
and then apply a POVM $X=\{X_{\alpha}\}$ on the original 
apparatus ${\cal H}_E$. It is shown that this strategy 
gives same amount of information with the original 
attack.  
The values obtained by the first measurement are
equally distributed, that is, each value $m$ is obtained 
with probability $\frac{1}{2^N}$. 
After obtaining a value $m$, 
the reduction of wave packet forces the state into
\begin{eqnarray*}
\rho^{i}_{m}:=\sum_{j}|E_{i\oplus m\ j\oplus m}\rangle \langle
E_{i\oplus m\ j\oplus m}|.
\end{eqnarray*}
The second measurement gives 
a probability 
\begin{eqnarray*}
p^s(\alpha|i,m)=\sum_j \langle E_{i\oplus m \ j\oplus m}
|X_{\alpha}|E_{i\oplus m\ j\oplus m}\rangle,
\end{eqnarray*}
from which it is easy to see that
$
p^s(\alpha|i,m)=p(\alpha|i\oplus m)
$
holds.
Thus by using 
conditional probability $p^s(\alpha,m|i)=\frac{1}{2^N}p(\alpha|i\oplus m)$,
mutual information can be computed to coincide with $I(A:E[X]|b)$. 
Taking a supremum over all the possible POVM over the full Hilbert space
${\cal H}_{E'}$ can make it larger and therefore the following 
inequality holds\cite{Biham},
\begin{eqnarray}
I(A:E|b)\leq I(A:E|b)_{sym},
\label{symineq}
\end{eqnarray}
where the right hand side is the optimal information gain 
by the symmetrized attack. 
\par
Now we can state our theorem.
\begin{theorem}
The following inequality holds:
\begin{eqnarray}
I(A:E|b)
\leq H(A\oplus B|\overline{b}),
\label{Ours}
\end{eqnarray}
where $H(\cdot)$ is the Shannon entropy.
That is, the information gain by Eve in the basis $b$ 
makes the outcome of measurement 
by Bob in the conjugate basis $\overline{b}$ random.
\end{theorem}
{\bf Proof:}
We can prove the theorem by 
first symmetrizing the attack
and next bounding Eve's information
gain by Holevo's inequality.
Thanks to (\ref{symineq}), it is sufficient to estimate 
the quantity $I(A,E|b)_{sym}$
for our purpose. 
Holevo's theorem\cite{Holevo}
bounds it from above as
\begin{eqnarray*}
I(A:E|b)_{sym}
&\leq& S\left(\frac{1}{2^N} \sum_i \rho^{i}_{Eve,sym}\right)
-\sum_{i}\frac{1}{2^N} S\left(\rho^i_{Eve,sym}\right)
\\
&=:&\chi\left(\{\rho^i_{Eve,sym}\}\right),
\end{eqnarray*}
where $S(\rho)$ is von Neumann entropy of a state $\rho$.
There exists a useful representation of this quantity $\chi$.
Consider another additional $N$-qubits Hilbert space ${\cal H}_R$ and 
a state over ${\cal H}_R\otimes {\cal H}_{E'}$,
\begin{eqnarray*}
\Theta:=\sum_{i}\frac{1}{2^N}|i\rangle \langle i| \otimes \rho^i_{Eve,sym}.
\end{eqnarray*}
Its quantum mutual entropy between ${\cal H}_R$ and 
${\cal H}_E$ is shown to coincide with 
the quantity $\chi\left(\{\rho^i_{Eve,sym}\}\right)$,
\begin{eqnarray*}
I(\Theta):=
S\left(\left.\Theta\right|_{E'}\right)
+S\left(\left.\Theta\right|_R \right)
-S\left(\Theta\right)
=\chi\left(\{\rho^i_{Eve,sym}\}\right),
\end{eqnarray*}
where $\left.\Theta\right|_{E'}$ is a restricted 
state to ${\cal H}_{E'}$ of $\Theta$ and $\left.\Theta\right|_R$ 
is defined in the same manner.
To estimate this quantity, we consider a 
purification of $\rho^{i}_{Eve,sym}$. 
Introduce another $N$-qubits system ${\cal H}_P$ 
and states over ${\cal H}_{E'} \otimes {\cal H}_P$ \cite{Biham},
$
|\varphi_i\rangle :=
\sum_j |E^s_{ij}\rangle \otimes |i\oplus j\rangle.
$
A state $\tilde{\Theta}$ over ${\cal H}_R \otimes {\cal H}_{E'} \otimes 
{\cal H}_P$ 
defined as
\begin{eqnarray*}
\tilde{\Theta}:=
\sum_i \frac{1}{2^N} |i\rangle \langle i| \otimes |\varphi_i \rangle
\langle \varphi_i|
\end{eqnarray*}
gives $\Theta$ if restricted to ${\cal H}_R \otimes {\cal H}_{E'}$. 
By using subadditivity for the entropy 
difference\cite{Thirring}, the mutual entropy $I(\tilde{\Theta})$ between 
${\cal H}_R$ and ${\cal H}_{E'}\otimes {\cal H}_P$ is shown to be larger than $I(\Theta)$. Therefore we estimate the quantity,
\begin{eqnarray*}
I(\tilde{\Theta}):=
S\left(\left.\tilde{\Theta}\right|_{E'P}\right)
+S\left(\left.\tilde{\Theta}\right|_R \right)
-S\left(\tilde{\Theta}\right).
\end{eqnarray*} 
Now we compute the restricted states over the subsystems,
\begin{eqnarray*}
\left.\tilde{\Theta}\right|_{R}&=&
\sum_{ij}\frac{1}{2^N}\langle E^s_{ij}|E^s_{ij}\rangle |i\rangle\langle i|
=\frac{1}{2^N}{\bf 1},
\\
\left.\tilde{\Theta}\right|_{E'P}
&=&\sum_i \frac{1}{2^N} |\varphi_i \rangle \langle \varphi_i|.
\end{eqnarray*}
The von Neumann entropy of $\left.\tilde{\Theta}\right|_{R}$ is 
$N$. \\
To compute the von Neumann entropy of $\tilde{\Theta}$ itself, 
we purify this by adding an additional $N$-qubits ${\cal H}_T$ and define 
a state over ${\cal H}_T \otimes {\cal H}_R \otimes
{\cal H}_{E'} \otimes {\cal H}_P$,
\begin{eqnarray*}
|\Psi\rangle :=\sum_i 
\sqrt{\frac{1}{2^N}}
|i\rangle \otimes |i\rangle \otimes |\varphi_i\rangle.
\end{eqnarray*}
Taking partial trace over ${\cal H}_R\otimes {\cal H}_{E'}\otimes {\cal H}_P$ 
leads 
\begin{eqnarray*}
\sum_i \frac{1}{2^N} \langle \varphi_i |\varphi_i\rangle |i\rangle\langle i|.
=\frac{1}{2^N} {\bf 1}
\end{eqnarray*}
whose entropy also is $N$.
Thus the mutual entropy is completely determined by 
$\left.\tilde{\Theta}\right|_{E'P}$ as
\begin{eqnarray*}
I(\tilde{\Theta})=S\left(\sum_i \frac{1}{2^N} |\varphi_i \rangle
 \langle \varphi_i|\right).
\end{eqnarray*}
Now let us calculate the von Neumann entropy of
 $\left.\tilde{\Theta}\right|_{E'P}$.
Again a purification 
using an additional $N$-qubits ${\cal H}_Z$ to 
${\cal H}_{E'}\otimes {\cal H}_P$ gives a state 
\begin{eqnarray*}
|\Phi\rangle 
:=\sum_i \sqrt{\frac{1}{2^N}}|i\rangle \otimes |\varphi_i\rangle
\end{eqnarray*} 
on ${\cal H}_Z \otimes {\cal H}_{E'}\otimes {\cal H}_P$. 
Its restriction to ${\cal H}_Z$ gives 
\begin{eqnarray*}
\sigma:=\frac{1}{2^N}
\sum_{ij}\sum_n \langle E^s_{j\ j\oplus u}|E^s_{i\ i\oplus u}\rangle |i\rangle
\langle j|
\end{eqnarray*}
whose entropy agrees with $I(\tilde{\Theta})$.
Let us consider its components with respect to the basis $\{|i\rangle\}$.
Since 
\begin{eqnarray*}
\sum_u \langle E^s_{j\ j\oplus u}|E^s_{i\ i\oplus u}\rangle
=
\frac{1}{2^N}
\sum_n \sum_u 
\langle E_{j\oplus u\ j\oplus n \oplus u}|
E_{i \oplus u\ i\oplus n \oplus u}\rangle
\end{eqnarray*}
holds, it depends upon only $i\oplus j$. 
We write it as
$f(i\oplus j)$ to represent $\sigma$ as
\begin{eqnarray*}
\sigma=\frac{1}{2^N}\sum_{ij}f(i\oplus j)|i\rangle \langle j|,
\end{eqnarray*} 
which can be diagonalized 
by an orthonormalized vectors 
\begin{eqnarray*}
|\mu_i\rangle:=\sqrt{\frac{1}{2^N}}
\sum_l (-1)^{i\cdot l}|l\rangle
\end{eqnarray*}
as
$
\sigma=
\sum_l \lambda_l |\mu_l\rangle \langle \mu_l|,
$
with $\lambda_l :=\frac{1}{2^N} \sum_t f(t)
(-1)^{t\cdot l}$.
The eigenvalue $\lambda_l$ is calculated as
\begin{eqnarray*}
\lambda_l &:=&\frac{1}{2^N} \sum_t f(t)
(-1)^{t\cdot l}
\\
&=&
\frac{1}{2^N} \sum_{t,n,v} \langle E_{v\ v\oplus n}
|E_{t\oplus v\ t\oplus v \oplus n}\rangle 
(-1)^{t\cdot l}
\\
&=&
\left(\frac{1}{2^N}\right)^2 
\sum_{t,n,v}\sum_{ij}\sum_{i'j'}
\langle \overline{E}_{ij}|\overline{E}_{i'j'}\rangle
\langle \overline{j}|v\oplus n\rangle \langle v|\overline{i}\rangle
\langle t\oplus v \oplus n|\overline{j'}\rangle
\langle \overline{i'}|t \oplus v\rangle 
(-1)^{t\cdot l}.
\end{eqnarray*}
Since we are treating mutually unbiased case,
\begin{eqnarray*}
\langle i| \overline{k} \rangle =\sqrt{\frac{1}{2^N}}(-1)^{i\cdot k}
\end{eqnarray*}
holds, where $i\cdot k:= \sum_{n=1}^N i_n k_n$. It leads
\begin{eqnarray*}
\lambda_l
&=&
\left(\frac{1}{2^N}\right)^4 
\sum \delta_{i\oplus j\oplus i' \oplus j',0}
\delta_{j\oplus j',0}
\delta_{j'\oplus i'\oplus l,0}
\langle \overline{E}_{ij}|\overline{E}_{i'j'}\rangle
\\
&=&
\frac{1}{2^N}\sum_i \langle \overline{E}_{i\ i\oplus l}|
\overline{E}_{i\ i\oplus l}\rangle 
\end{eqnarray*}
which is nothing but $p(B=A\oplus l|\overline{b})$ introduced in 
(\ref{defofP1}). 
Finally we obtain the following inequality,
\begin{eqnarray*}
I(A:E|b)
\leq H(A\oplus B|\overline{b}).
\end{eqnarray*}
\hfill Q.E.D.
\section{Discussions}
Below we discuss the implication of our 
theorem by comparing it with the former one.
Since the right hand side of our inequality is 
determined by $\{p(B=A\oplus c|\overline{b})\}$,
it can be reduced to a form which includes only the term
$\sum_{c\neq 0} p(B=A\oplus c|\overline{b})$.
\begin{corollary}\cite{Hayashi}
The following inequality between 
the information gain by Eve and the error probability in 
Bob's outcome holds:
\begin{eqnarray*}
I(A:E|b)
\leq -\delta \log \delta -(1-\delta)\log(1-\delta)
+ N \delta,
\end{eqnarray*}
where $\delta:=\sum_{c\neq 0}p(B=A\oplus c|\overline{b})$.
\end{corollary}
{\bf Proof:}
\\
Under the constraint $\delta=\sum_{c\neq 0}p(B=A\oplus c|\overline{b})$ 
for fixed $\delta$, the distribution which makes 
the Shannon entropy $H(A\oplus  B|\overline{b})$ maximum is 
$p(B=A|\overline{b})=1-\delta$ and $p(B=A\oplus c|\overline{b})=
\frac{\delta}{2^N-1}$ for all $c \neq 0$. 
It gives 
\begin{eqnarray*}
H(A\oplus B|\overline{b})
=-\delta \log \delta -(1-\delta)
\log \delta +\delta \log (2^{N}-1)
\end{eqnarray*}
and ends the proof.
\hfill Q.E.D.
\par
For a fixed error probability $\delta=\sum_{c\neq 0}
p(B=A\oplus c|\overline{b})$, for 
sufficiently large $N$, 
the term $N \delta$ becomes dominant
in the right hand side of the above equation. 
Thus our inequality becomes tighter than (\ref{Boykins}) in 
such a case. 
\par
Finally we present a situation which shows a drastic 
difference between the two inequalities.
Suppose that Eve employs the following ``attack":
Eve does not make the qubits sent by Alice interact with 
any apparatus, but she just converts the each value.
That is, for each qubit, Eve performs a
unitary operation $|i\rangle \mapsto (-1)^i |i\rangle\ (i=0,1)$.
One can easily see that also for the conjugate basis
this operation works as conversion. 
In this case the error probability $\delta$ becomes 
$1$. Thus if we employ 
the inequality (\ref{Boykins}), 
it is impossible to rule out 
the possibility of Eve's information gain.
On the other hand, since the error in Bob's outcome 
is deterministic, the right hand side of 
(\ref{Ours}) vanishes. Thus our theorem 
can convince us that there is no 
information gain by Eve.
\par
In this paper we showed a novel version of 
information-disturbance theorems. According to 
our theorem, one can see that 
the information gain by Eve induces randomness to 
Bob's outcome in the conjugate basis. 
The both sides of the inequality are 
expressed in terms of entropy and thus seems to be 
natural. For large $N$ case, in which we are 
usually interested in, our inequality gives tighter bound 
than the previously proposed ones. 
Moreover, our theorem can rule out the case
when Eve just turns over the qubits and gains no information. 
Our theorem, as previous one, also relies upon 
the assumptions of fair probability of the random variable $A$ 
and mutually unbiasedness between $b$ and $\overline{b}$. 
It will be very interesting and crucial to generalize 
the theorem to more general setting\cite{della}.


\begin{thebibliography}{9}
\bibitem{Note}
In \cite{Boykin}, they treated 
the case of $D$ (not restricted with $D=2$) dimensional 
Hilbert space.
\end{thebibliography}
\end{document}